\documentclass[prb,twocolumn,preprintnumbers,amsmath,aps]{revtex4}
\usepackage{graphicx,hyperref}
\usepackage{bm}
\usepackage{subfigure}

\begin{document}

\title{An assessment of the concept of fragility}

\author{Gilles Tarjus} \email{tarjus@lptl.jussieu.fr}
\affiliation{LPTMC, CNRS-UMR 7600, Universit\'e Pierre et Marie Curie,
bo\^ite 121, 4 Pl. Jussieu, 75252 Paris Cedex 05, France}

\author{Christiane Alba-Simionesco} \email{christiane.alba-simionesco@cea.fr}
\affiliation{  LLB, UMR 12, CEA-CNRS, 91191 Saclay, France}

\begin{abstract}
The concept of ``fragility'', which was introduced to characterize the degree of 
super-Arrhenius temperature dependence of the relaxation time and transport coefficients, 
has since been taken as a key quantity that seems to correlate with other properties of 
glass-forming liquids and polymers. With the goal of assessing the usefulness and the robustness of 
the concept, we address here several questions: How to best quantify fragility? How 
significant are the observed differences in fragility? Is fragility connected to 
``cooperativity'' and collective behavior?
\\

\textit{Proceedings of the Symposium on ``Fragility of Glass-forming Liquids'', A. L. Greer, K. F. Kelton, and S. Sastry eds., 
TRIPS 13 (New Dehli, 2014)}

 \end{abstract}

\maketitle

\section{Introduction}

The concept of ``fragility'', introduced by Angell \cite{angell85}, characterizes how quickly transport coefficients and relaxation times increase as 
one cools a glass-forming system. It focuses on a \textit{generic}, striking, property of the dynamical slowdown in most glass-forming 
molecular liquids and polymers, namely the fact that the temperature dependence of the dynamical properties is stronger than an 
Arrhenius one. Quantifying fragility amounts to characterizing the degree of super-Arrhenius behavior, which is material \textit{specific}. 

In trying to meaningfully rationalize the phenomenological observations on glass formation, correlations have then been empirically 
investigated for a variety of supercooled liquids and polymers between measures of fragility and other 
material-specific quantities, be they thermodynamic, such as the amplitude of the heat-capacity jump at the glass transition or 
the steepness of the decrease of the configurational or excess entropy with temperature, or dynamic, such as the ``stretching'' of 
the time-dependent $\alpha$-relaxation functions, the intensity of the Boson peak or other vibrational or short-time properties.

For useful as it has been in the search for developing a full understanding and a theory of glass formation, the 
concept of fragility has also weaknesses. It may therefore be interesting, some 30 years after its introduction, to assess its 
relevance, both at an operational and at a fundamental level. 
Among the questions that can be raised about fragility, we would like to discuss the three following ones: How to best quantify 
fragility? How significant are the observed differences in fragility? Is fragility connected to ``cooperativity'' and collective behavior?
The following is a brief introduction, and some issues are illustrated in the accompanying reprints. Needless to say that we are happy 
to dedicate it to our friend Austen Angell for his inspiring contribution to the field.

\section{How to best quantify fragility?}

Ideally, one would like to associate the fragility of a given glass-former with a well-defined index that captures the \textit{intrinsic} 
characteristic of the super-Arrhenius slowdown. Fragility is easily visualized, {\it e.g.} on the celebrated Angell plot \cite{angell85}, 
yet a precise quantitative measurement is not straightforward. The uncertainty in the value extracted for a given glass-former 
may then cast doubt on the robustness of the correlations discussed above.

Beyond the significant error bars that are found due to manipulations and data uncertainty, one first faces the problem of defining an 
appropriate reference temperature in order to rescale the behavior of the various glass-forming 
systems on a dimensionless temperature axis. The choice of this reference temperature is neither obvious, as there are no observed 
singularities in the behavior of the glass-formers, nor benign. The common practice is to consider a ``low'' reference temperature, usually 
the empirically determined glass transition temperature $T_g$ (as on the Angell plot). The canonical measure of fragility is then the 
``steepness index'' at $T_g$ \cite{steepness}: $m=\partial \log_{10}[\tau(T)/\tau_{\infty}]/\partial(T_g/T)\vert_{T_g}$. 

Choosing $T_g$, or a related low reference temperature, has several shortcomings. First (but not necessarily the most important), it does not allow 
a practical comparison between liquid models studied by computer simulation (that never span more than 5 orders of magnitude 
in time) and experiments on molecular liquids that are able to probe much slower dynamics. Second, $T_g$ and related 
temperatures  are only defined through a long but arbitrary time scale. As a result, they do not carry any particular physical content. 
Characterizing fragility by a parameter calculated at $T_g$ may then include spurious effects. If the temperature evolution of 
the relaxation time (or viscosity) of a glass-forming liquid can be described as being Arrhenius-like at high temperature,
\begin{equation}
\label{eq_HS_mastercurve}
\tau(T)=\tau_{\infty}\, {\rm exp}\left (\frac{E_{\infty}}{T}\right )\, ,
\end{equation}
as usually found to a good approximation \cite{kivelson,rossler}, and super-Arrhenius at low temperature, 
\begin{equation}
\label{eq_HS_mastercurve}
\tau(T)=\tau_{\infty}\, {\rm exp}\left (\frac{E(T)}{T}\right )
\end{equation}
with $E(T)$ growing as $T$ decreases, which precisely corresponds to the notion of fragility, one realizes that the properties of the 
relaxation at high temperature may play a role in the behavior observed at $T_g$. Clearly, a large effective activation energy 
$E_{\infty}$ at high temperature leads by itself to a rapid slowdown of the dynamics that influences the value of the steepness 
index at $T_g$. Fragility \textit{per se} needs then to be disentangled from this high-$T$ slowdown \cite{extension}. This problem 
could be circumvented by ``subtracting'' the effect of the high-$T$ 
Arrhenius dependence, {\it e.g.} by considering the function $(E(T)-E_{\infty})/E_{\infty}$. The downside however is that one needs 
additional manipulations of the data and that the premisses of this operation may not be unanimously accepted.

An alternative procedure is to choose a ``high'' reference temperature, taken as a crossover or onset temperature at which 
super-Arrhenius behavior starts to be detectable. The choice of such a high-$T$ crossover point has the merit of allowing a direct comparison 
of simulation and laboratory data \cite{fragilityLJ,coslovich} and also easily leads  to a subtraction of the high-$T$ effect (see above) in order to define 
an intrinsic fragility that do not depend on an arbitrary time scale \cite{kivelson}. However, finding a robust operational way for defining the 
crossover is far from trivial. 

Another potential problem in the definition of fragility is that the latter involves a variation with temperature that {\it a priori} depends on the 
thermodynamic path chosen. The fragility index is usually defined at constant pressure. Doing so, one includes 
in the fragility measure not only the intrinsic effect of temperature but also the influence of the increasing density. To get around this, one 
should use a constant-density (``isochoric'') fragility in place of the standard ``isobaric'' one. Experimental data however are not collected 
under isochoric conditions and this makes the general use of the isochoric fragility more difficult.

A major simplification comes from the recently unveiled existence of an approximate empirical scaling for glass-forming liquids and polymers, 
of the form $\tau(\rho,T)/\tau_{\infty}={\rm exp}[\mathcal F(e(\rho)/T)]$ with $\mathcal F$ a material-dependent but state-point independent 
scaling function \cite{alba_scaling,roland_scaling,dreyfus_scaling}.  The direct consequence of the scaling is that the isochoric fragility, no 
matter the operational procedure by which it is measured, does not depend on density:  see reprint 1 \cite{reprint1}. (The scaling being approximate 
for molecular liquids and polymers, except for liquid models with inverse power-law repulsive interactions, this independence is also 
approximate.) The isochoric fragility is thus a better starting point for an {\it intrinsic} characterization 
of the super-Arrhenius $T$-dependence of a given glass-forming liquid or polymer: see reprint 1.  
The issues discussed in the first part of this section of course still remain.

\section{How significant are the observed differences in fragility?}

The breadth of fragilities between a strong glass-former like silica and fragile liquids or polymers seems impressive: it corresponds 
to isobaric steepness indices at $T_g$ (and atmospheric pressure) ranging from about 20 (silica) to 80-100 for fragile molecular 
liquid such as ortho-terphenyl or toluene and 150 and more for some polymeric systems. Such a large spectrum appears to call for 
an explanation unveiling the physical sources of fragile behavior and the origin of the observed differences between materials.

We have already discussed possible ``spurious'' effects entering into the isobaric steepness indices 
at $T_g$: the more or less important influence of the high-$T$ dynamical behavior and of the associated effective activation 
energy scale on the one hand and the role of density, which increases with decreasing temperature at constant pressure,  
on the other. These two effects, which involve the strength of the ``bare'' activation energy and thermodynamic parameters such as 
the coefficient of thermal expansivity, vary from system to system with, most likely, very little connection with the generic properties of 
the (fragile) viscous slowing down. In a similar vein, it has been argued that the very high fragility reached by polymers compared 
to molecular liquids come from features specifically associated with the chain structure of the former: see reprint 2 \cite{reprint2}. 
Some differences in fragility therefore appear to be rationalizable without invoking anything fundamental concerning the glass transition.

As far as differences in fragility are concerned, it is instructive to compare the behavior of glass-forming liquids to that of the glassy 
systems considered in the context of ``jamming'' phenomena, such as foams and emulsions \cite{durian}. Simple models for the latter 
consist of spherical particles interacting through pairwise truncated repulsive potentials, {\it e.g.} $v( r)=\epsilon(1-r/\sigma)^{\alpha}$ 
for $r<\sigma$ and is zero 
otherwise (the exponent $\alpha$ is typically taken to be $2$ or $3/2$) \cite{durian}. When studied at low temperature via computer simulations, 
it has been found that the isochoric fragility of the systems strongly varies with density, up to an order of magnitude 
(but possibly more) \cite{berthier-witten}. The reason in this case comes from the fact that such systems can be considered 
as effective hard-sphere models \cite{berthier-witten,liu-nagel}.

The reasoning goes as follows. For the sake of simplicity, we consider a (hypothetical) one-component system and describe the 
relaxation of the hard-sphere fluid by a free-volume-like formula
\begin{equation}
\label{eq_HS_relaxation}
\tau(\eta)\propto {\rm exp}\left (\frac{\eta_0}{\eta_0-\eta}\right )
\end{equation}
where $\eta=(\pi/6)\rho\sigma^3$ is the packing fraction, $\sigma$ the hard-sphere diameter, and $\eta_0$ a random-close packing 
at which the pressure diverges (one can define such a singularity for hard-core objects). At low $T$, the structural  and dynamical 
properties of systems interacting with truncated repulsive interactions can be very well described as those of a hard-sphere model 
with an effective hard-core diameter, {\it i.e.} an effective packing fraction $\eta_{eff}(\eta,T) \sim \eta/(1+a\sqrt{T}) \leq \eta$. For 
$T>0$, one then has
\begin{equation}
\label{eq_HS_mastercurve}
\tau(\rho,T)=\tau_{\infty} {\rm exp}\left (\frac{\eta_0}{\eta_0-\eta_{eff}(\eta,T)}\right )
\end{equation}
where $\tau_{\infty} \propto 1/\sqrt T$.
As a result, when quantifying for instance the isochoric fragility by the steepness index $m^*_{\eta}$ at a chosen temperature $T^*$ 
where $\tau/\tau_{\infty}$ takes a large given value $e^{K^*}$ (typical, say, of the calorimetric glass transition), one obtains for $\eta 
\gtrsim \eta_0$
\begin{equation}
\label{eq_HS_fragility}
\begin{aligned}
m^*_{\eta}(\eta)&= \frac{\partial \log_{10}[\tau(\eta,T)/\tau_{\infty}]}{\partial(T^*/T)}\bigg \vert_{T=T^*(\eta)}\\&
\simeq \frac{(K^*)^2}{4.6}\left (1-\frac{\eta_0}{\eta} +\mathcal O(1/K^*)\right ) \,.
\end{aligned}
\end{equation}
As $K^*$ can be large ($\sim 37$ at the glass transition), even a moderate change of the packing fraction in the vicinity of $\eta_0$  
can generate a large variation of the isochoric fragility, as indeed observed.

The above behavior, which is shared by all jamming system with truncated repulsive interactions when considered at nonzero temperature, 
is at odds with what is found is glass-forming liquids and polymers, where as discussed in the previous section the isochoric fragility is 
essentially independent of density: see reprint 3 \cite{reprint3}. The variation, or lack thereof, of the isochoric fragility is here a clearcut feature 
that distinguishes jamming phenomenology from glass-forming liquids and polymers \cite{berthier-tarjus}.

\section{Is fragility connected to ``cooperativity'' and collective behavior?}

Fragility, {\it i.e.} the fact that the dynamical slowdown has stronger than Arrhenius temperature dependence, is suggestive of 
growing ``cooperative'' behavior (as temperature decreases) and has often been taken as a central property of glass-formers 
for this reason. ``Cooperativity'' in the context of thermally activated dynamics means that degrees of freedom must conspire 
to make the relaxation possible (or faster than by other means). In consequence, the effective barrier to relaxation is determined 
by the minimum number of degrees of freedom that are cooperatively involved. This idea is at the core of many theoretical 
approaches of the glass transition, including the Adam-Gibbs notion of  ``cooperatively rearranging regions'' \cite{adam-gibbs65}.

Beyond the Adam-Gibbs somewhat heuristic picture, one can try to relate fragile behavior to the existence of a characteristic length 
scale. The main ideas behind such a relation are that (i) if a system has a finite correlation length it can be divided into independent 
subsystems of size larger than (but of the order of) this correlation length and (ii) a finite-size system has a finite relaxation time whose 
magnitude can be related to the its size. In the absence of any obvious form of order in glass-forming liquids and polymers (and 
facing the experimental fact that structural pair correlations are of little or no help), the relevant (static) length should describe 
how far a condition at the boundary can influence the interior of a (sub)system, a notion 
that is captured by the so-called ``point-to-set'' correlation 
lengths. Let's $\xi(T)$ denote the largest point-to-set correlation length at temperature $T$ and assume that relaxation even in 
the high-$T$ non-cooperative regime proceeds by thermal activation (as empirically found). Then, a finite (independent) subsystem 
of linear length of the order of $\xi(T)$ is expected to relax to equilibrium in a typical time 
\begin{equation}
\label{eq_time_PTS}
\log[\tau(T)/\tau_{\infty}]\simeq \frac{A}{T}\xi(T)^{\psi}
\end{equation}
with $\psi \leq d$. (A true upper bound, which can be made rigorous under some conditions \cite{montanari06}, is obtained when 
$\psi=d$, as the most costly barrier to relaxation involves the full volume or the total number of particles). 
Any measure of fragility then leads to the conclusion that super-Arrhenius behavior implies that the length $\xi(T)$, which  
quantifies cooperativity in a precise manner, increases with decreasing temperature.

In practice, because the growing correlation length is expected to appear at some power larger than 1 in the argument of an exponential [see 
Eq. (\ref{eq_time_PTS})], a small increase of this length is enough to generate changes of the relaxation time by orders of magnitude. 
Matter is even worse as the point-to-set length can (so far?) only be measured in 
computer simulations where a limited increase of the relaxation time (and an even more limited one of the static length) is probed. 
As for experiments, they only give access to some multi-point dynamical susceptibilities, hence to a rough estimate of a different type of 
length that characterizes the correlations in the dynamics associated with the phenomenon of dynamical heterogeneities.  This then 
leaves room for a variety of alternative views and hot debates.

In the above discussion, we have considered the evidence provided by a generic fragile behavior concerning the growing 
cooperative character of the relaxation and we have recalled arguments in favor of a positive conclusion (to be however taken with a grain 
of salt). Accordingly, a strong, Arrhenius-like, system would have essentially non-cooperative dynamics. One might further ask if there is a 
relation between the magnitude of the fragility in a glass-former and its more or less cooperative dynamical behavior?  Unfortunately, 
in the absence of a well-defined theoretical framework, it does not seem obvious to make sense of the question.

\end{document}